\documentclass[fleqn,10pt]{wlscirep}
\title{Modelling realistic microgels in an explicit solvent}
\let\vec\mathbf
\author[1,2,*]{F. Camerin}
\author[1,3]{N. Gnan}
\author[1,3]{L. Rovigatti}
\author[1,3,*]{E. Zaccarelli}
\affil[1]{CNR-ISC, Uos Sapienza, Piazzale A. Moro, 2, 00185, Roma, Italy}
\affil[2]{Dipartimento di Scienze di Base e Applicate per l'Ingegneria, Sapienza Universit\`a di Roma, via A. Scarpa, 14, 00161, Roma, Italy}
\affil[3]{Dipartimento di Fisica, Sapienza Universit\`a di Roma, Piazzale A. Moro, 2, 00185, Roma, Italy}

\affil[*]{fabrizio.camerin@uniroma1.it, emanuela.zaccarelli@cnr.it}


\begin{abstract}
Thermoresponsive microgels are polymeric colloidal networks that can change their size in response to a temperature variation.
This peculiar feature is driven by the nature of the solvent-polymer interactions, which triggers the so-called volume phase transition from a swollen to a collapsed state above a characteristic temperature. Recently, an advanced modelling protocol to assemble realistic, disordered microgels has been shown to reproduce experimental swelling behavior and form factors. In the original framework, the solvent was taken into account in an implicit way, condensing solvent-polymer interactions in an effective attraction between monomers. To go one step further, in this work we perform simulations of realistic microgels in an explicit solvent. We identify a suitable model which fully captures the main features of the implicit model and further provides information on the solvent uptake by the interior of the microgel network and on its role in the collapse kinetics. These results pave the way for addressing problems where solvent effects are dominant, such as the case of microgels at liquid-liquid interfaces.
\end{abstract}
\begin{document}

\flushbottom
\maketitle

\thispagestyle{empty}

\section*{Introduction}
\bigskip

In recent years microgels --- colloidal-scale polymer networks --- have emerged as a popular model system in condensed matter physics\cite{yunker2014physics} thanks to their colloid/polymer duality\cite{lyon2012polymer}. The combination of colloidal properties and responsiveness to external stimuli is the key for their appeal for both applications and fundamental science\cite{fernandez2011microgel}. Among microgels, the most widely studied are those based on Poly(N-isopropylacrylamide) (PNIPAM), a thermoresponsive polymer able to swell and deswell reversibly as a result of temperature changes. When PNIPAM chains are crosslinked with bisacrylamide (BIS), microgel particles can be prepared in a range of sizes of $10-1000$~nm by standard synthesis methods \cite{pelton2000temperature}, and even reach much larger scale (up to $100\,\mu$m) with microfluidic techniques~\cite{seiffert2013microgel}. These particles undergo a Volume Phase Transition (VPT) in water at a temperature of $\approx305K$, from a swollen state at low temperatures to a collapsed one at high temperatures. This swelling-deswelling transition is fully reversible and can be exploited to tune the size of the particles {\it in situ}. The VPT is completely controlled by the polymer-solvent interactions, echoing the coil-to-globule transition of linear PNIPAM chains in water\cite{tavagnacco2018}. As a matter of fact, the role of water is highly relevant, as the VPT originates from changes in the hydrophilic/hydrophobic character of the interactions of the polymer with the solvent upon temperature variations.

Experimental work on microgels has enormously increased in the last couple of decades, and a comparison of experimental data with theory has been possible thanks to the use of the classical Flory-Rehner theory of swelling\cite{saunders1999microgel}. On the other hand, microgel simulations have been less abundant due to the complex, multi-scale nature of the particles. So far, most efforts have relied on the use of unrealistic networks, often based on ordered, diamond-like topologies, in which all strands are of equal length\cite{jha2011study,ghavami2016internal,kobayashi2017polymer,Ahualli2017,nikolov2018mesoscale}. Only a few of these approaches have explicitly considered the role of the solvent\cite{ghavami2017solvent,kobayashi2014structure,nikolov2018mesoscale}.

Recently, we have introduced a novel method to synthesize realistic microgel particles {\it in silico} through the assembly of fully-bonded, disordered networks with arbitrary topology\cite{gnan2017,rovigatti2018internal}. In this approach we initially consider the self-assembly of a mixture of patchy particles, respectively bivalent and tetravalent, to mimic monomers and crosslinkers. To retain a spherically-shaped network, the mixture is confined within a sphere of a given radius. Fully-bonded configurations are obtained by introducing a swapping mechanism that makes it possible to equilibrate the system even at the strong attractions required to maximize the bonding. 
In this protocol there are two parameters controlling the topology of the resulting network: the concentration of crosslinkers and the confinement radius. Thus, more compact and homogeneous networks are obtained in presence of a large number of crosslinkers and/or for a tight confinement, while looser and more heterogeneous microgels can be produced with a smaller amount of crosslinkers and a very weak confinement. A thorough discussion on how the internal structure of the microgels depends on these parameters can be found in Refs.\cite{gnan2017,rovigatti2018internal}.

Once the network is assembled, we replace the patchy (reversible) interactions with permanent bonds by adopting the classical Kremer-Grest bead-spring model for polymers\cite{grest1986molecular} to preserve the network topology throughout the course of the simulation.
In order to reproduce the swelling behavior, it is possible to incorporate in the model an attractive potential that has been shown to capture the variation in polymer-water interactions upon changing temperature. With this approach, the solvent is implicit and the solvophobic potential accounts for it within the thermodynamic properties of the system in an effective way. This implicit solvent model was shown to be able to faithfully reproduce swelling data of individual microgels measured with Dynamic Light Scattering experiments \cite{gnan2017}.
Even though the use of an explicit versus an implicit solvent model\cite{pham2009implicit,spaeth2011comparison} should give identical results in terms of equilibrium properties, there are a number of features that cannot be correctly captured and/or described by an implicit model. In particular, the kinetics of swelling and deswelling will depend on the presence of the solvent and on how it is modelled. Besides that, there are situations of fundamental and practical interest in which an explicit solvent will dramatically affect the picture. For instance, to model a system at a liquid-liquid interface, it is necessary to take into account the presence of the two different media in order to capture effects related to the surface tension \cite{neyt2014quantitative,rumyantsev2016polymer}.

In order to be able to handle these situations, here we take the implicit-solvent model of Refs.\cite{gnan2017, rovigatti2018internal} and extend it by developing an explicit solvent description that accurately predicts the swelling behavior of microgel particles. We use the swelling properties exhibited by the implicit solvent model, which has been shown to faithfully reproduce the experimental results, as reference data to calibrate the explicit-solvent parametrisation. By comparing the swelling ratio as a function of temperature and the microgel density profile and form factor with and without solvent, we are able to discriminate among different solvent models and choose the explicit description that works best. In particular, we intend to model a generic solvent that ensures that the key properties of microgel colloids are accurately reproduced rather than to provide a systematic and exhaustive study on the influence of the system parameters on the properties of the particle. We further test the robustness of our approach by repeating the analysis for microgels generated with different topologies and confinement radii. Once established our explicit model, we first look at the arrangement of the solvent inside the microgel across the volume phase transition, and then study the kinetics of the deswelling. Overall, our results open up the possibility to obtain more and more realistic descriptions of microgels, thanks to which it will be possible to tackle exciting problems in which the explicit role of the solvent plays a crucial role \cite{rumyantsev2016polymer,isa2017two,scheidegger2017compression,brugger2010interfacial}. 

\section*{Results}
\subsection*{Swelling behavior}
\begin{figure}[h!]
\centering
\includegraphics[scale=0.50]{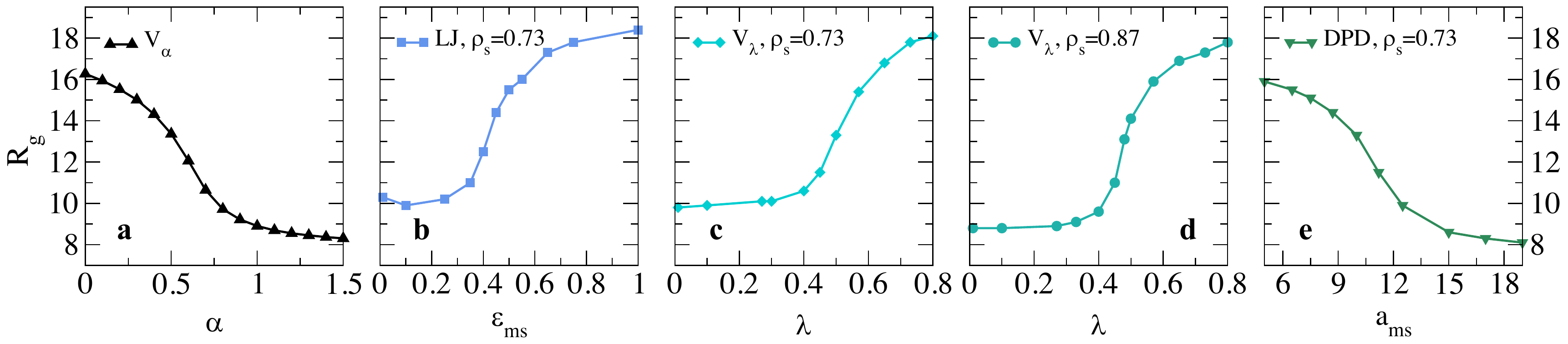}
\caption{\small \textbf{Microgel swelling curves.} Radius of gyration $R_g$ across the VPT transition for (a) the implicit model, $V_{\alpha}$; (b) the explicit LJ solvent with LJ monomer-solvent interactions at a solvent density $\rho_s=0.729$; (c,d) explicit solvent with $V_\lambda$ monomer-solvent interactions at $\rho_s=0.729$ and $\rho_s=0.875$, respectively; (e) DPD simulations where the microgel is modeled as a bead-spring polymer network. All curves report the gyration radius $R_g$ as a function of the parameter controlling the solvophobic interactions in each model: (a) $\alpha$, (b) $\epsilon_{\rm ms}$, (c-d) $\lambda$ and (e) $a_{\rm ms}$. }
\label{fig:cfr_swelling}
\end{figure}

\begin{figure}[hb!]
\centering
\includegraphics[scale=0.4]{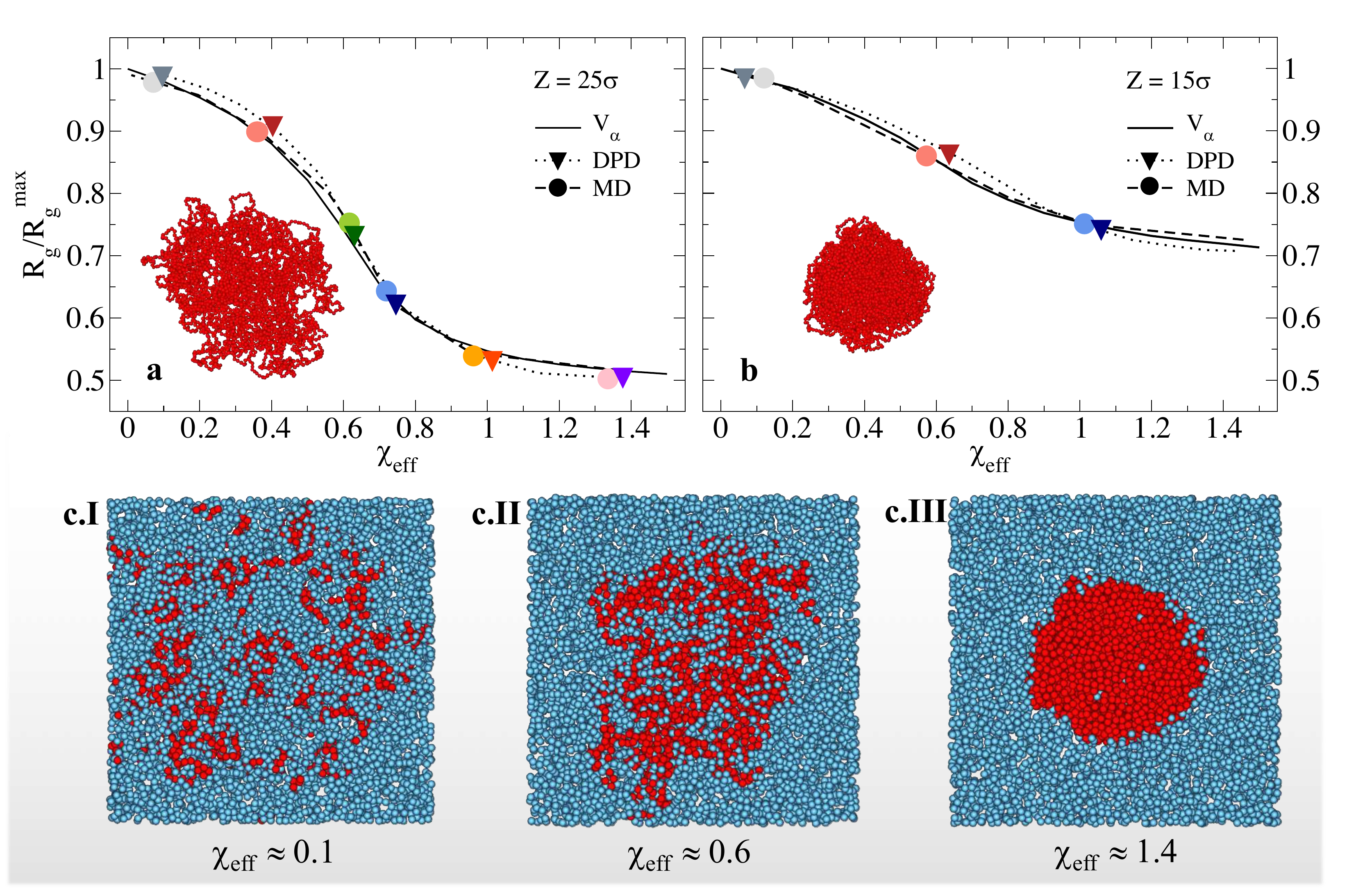}
\caption{\small\textbf{Effect of microgel topology and solvent arrangement.}  Swelling curves for the implicit- (full line) and explicit-solvent models that best reproduce the swelling behavior, namely MD simulations with $V_{\lambda}$ at $\rho=0.87$ (dashed lines) and DPD simulations (dotted lines) for (a) a loose microgel ($Z=25\sigma$) and (b) a more compact microgel ($Z=15\sigma$). Corresponding microgel snaphots are also shown.  Symbols refer to state points in explicit solvent simulations (MD: circles, DPD: triangles) for which further analysis is provided in the next sections, whereas similar colors/shapes refer to similar swelling degrees between the two explicit solvent models. Panels (c.I-c.III) display a central slab of the simulation box for three different values of $\chi_{\rm eff}$, respectively corresponding to the swollen state (c.I), a state very close to the VPT (c.II) and the collapsed state (c.III).  The arrangement of solvent (blue spheres) within/around the polymer network (red spheres) depends on $\chi_{\rm eff}$. For visual clarity, only half of the solvent particles are shown.}
\label{fig:swellingcolori}
\end{figure}
We start by discussing the swelling behavior of microgels in the presence of an explicit solvent as compared to the reference case of the implicit model $V_{\alpha}$, Fig. \ref{fig:cfr_swelling}(a) (see Methods), discussed in Ref.\cite{gnan2017}. To this aim, we perform simulations of an individual microgel assembled with a rather loose topology (using a confining radius $Z=25\sigma$) in different solvents. In particular, we make a comparison between ``atomistic'' and coarse-grained solvent representations by employing Molecular Dynamics (MD) and Dissipative Particle Dynamics (DPD) simulations, respectively (see Methods). In the former type of approach, we first need to adjust the solvent-solvent interactions, for which the most natural choice is to use a Lennard-Jones (LJ) potential.
Next, we address the choice of the monomer-solvent (ms) interactions: these are crucial to describe the swelling transition, because they control the contraction or extension of the polymers chains in the solvent environment. To discriminate between different models and identify the best possible candidate, we explore multiple ms potentials and compare them to the implicit solvent case. The choice of the solvent density allows to tune the pressure exerted by the solvent on the polymer network thus determining the swelling range of the microgel particle, as discussed below.

Similarly to solvent-solvent interactions, a straightforward choice for the monomer-solvent ones is the LJ potential\cite{schwenke2014conformations} where, by varying the energy minimum $\epsilon_{\rm ms}$, we control the polymer-solvent affinity. In this way, we obtain the swelling curve reported in Figure \ref{fig:cfr_swelling}(b), where the radius of gyration of the microgel $R_g$ is shown as a function of $\epsilon_{\rm ms}$: by decreasing this parameter (with respect to solvent-solvent interaction, which sets the energy scale), the polymer-solvent interactions are less favoured than solvent-solvent ones, giving rise to a reduction of the microgel size. However, an unphysical increase of $R_g$ is observed for $\epsilon_{\rm ms}\rightarrow 0$: under this condition, both terms in the LJ potential go to zero, \textit{i.e.} the microgel feels neither attraction nor repulsion with the solvent. Consequently, the network relaxes as the external pressure on the polymer network vanishes, and the microgel swells again, maximizing its configurational entropy.

Such behaviour clearly indicates the unsuitability of the LJ potential to mimic the solvent-monomer interactions. Consequently, the next step is to use a potential in which the attractive term can be tuned arbitrarily without affecting the short-range repulsion. To this aim, we adopt the $V_{\lambda}$ model, defined in Eq.~\eqref{lambda}, where the repulsion remains unchanged while the attractive contribution, controlled by the parameter $\lambda$, is varied.
The swelling behavior of the microgel obtained with this model is reported in Fig.~\ref{fig:cfr_swelling}(c,d) for two representative solvent densities. The swollen-to-collapsed transition is well reproduced in both cases.

So far, we have assessed the ``atomistic'' type of solvent. We further examine the possibility to use a coarse-grained solvent by means of DPD simulations, which correctly reproduce hydrodynamic interactions at long times\cite{groot1997dissipative}. In order to establish a meaningful comparison with the implicit solvent case and avoid unphysical crossing of the chains, we retain the bead-spring model for monomer-monomer interactions and we apply the DPD treatment only to monomer-solvent and solvent-solvent interactions.

The results of DPD simulations, for the parameters specified in Methods, are reported in Fig.~\ref{fig:cfr_swelling}(e). In this case, the VPT transition is modulated by the monomer-solvent repulsion quantified by the parameter $a_{\rm ms}$ in Eq.~\eqref{dpd_fc}: for small values of $a_{\rm ms}$ the microgel is swollen, while it contracts when $a_{\rm ms}$ increases. We notice that $R_g$ is systematically larger at comparable swelling for MD-solvents than for DPD results, which, on the other hand, quantitatively reproduce the values obtained in the implicit solvent description. This is due to the softness of the DPD interactions which, contrarily to the MD treatment, do not introduce significant solvent-monomer excluded volume effects, thereby not affecting the microgel size.

In order to establish a correspondence between different models, we rescale the explicit solvent data onto the implicit one, $V_\alpha$ where $\alpha$ is the solvophobic parameter (see Methods). Figure~\ref{fig:swellingcolori}(a) shows the normalized $R_g/R_g^{max}$, where $R_g^{max}$ is the value of the gyration radius at maximum swelling, as a function of the effective swelling parameter $\chi_{\rm eff}$. The latter corresponds to the solvophobic parameter $\alpha$ of the implicit solvent simulations. We report the comparison for the two cases where the agreement is found to be fully satisfactory for all $\chi_{\rm eff}$, namely the DPD and MD $V_{\lambda}$ models. Of the latter, we consider only the case with the highest solvent density, $\rho=0.87$, since deviations with respect to the implicit solvent case are observed with lower densities: the swelling range of the microgel would be shortened, as can be observed in Figure \ref{fig:cfr_swelling}(c).
Thus, it appears that, while $V_{\lambda}$ is definitely superior to the simple LJ potential to model the VPT of the microgel, the density of the solvent particles is a key parameter in tuning the details of the transition: a lower density will have a smaller effect on the microgel, resulting in a more limited contraction with respect to the implicit solvent model. From now on we will discard the LJ potential and we will refer to MD simulations as those performed with the $V_{\lambda}$ interaction.  A similar effect can be obtained in DPD simulations by changing the cutoff radius and the interaction parameters of the conservative force, which represents the length scale in DPD and the size of the solvent beads (see Methods).

To verify the robustness of our protocol, we now repeat the above analysis on a microgel configuration assembled with a smaller confinement radius, $Z=15\sigma$. Fig.~\ref{fig:swellingcolori}(b) reports the swelling behaviour of the more compact microgel for the DPD and MD  models at the optimal solvent density identified above. Together with the data, we also report snapshots of the two microgels (insets) in their maximally swollen state, showcasing the very different topology of the networks. The good agreement between the rescaled swelling curves for both studied microgel configurations allows us to conclude that the developed models are robust and both can faithfully reproduce the swelling behavior observed with the implicit model\cite{gnan2017}. Fig.~\ref{fig:swellingcolori}(c.I-c.III) further highlights the arrangement of solvent particles inside the microgel for MD simulations at different values of $\chi_{\rm eff}$ across the VPT. The microgel remains very permeable to the solvent even close to the transition temperature, finally expelling it only in the fully compact state. 
In the next sections we focus on MD and DPD to study the effects of the solvent on the microgel structural features and on the kinetics of the volume phase transition. 

\subsection*{Structural features of a loose microgel in an explicit solvent}
\begin{figure}[b!]
\centering
\includegraphics[scale=0.45]{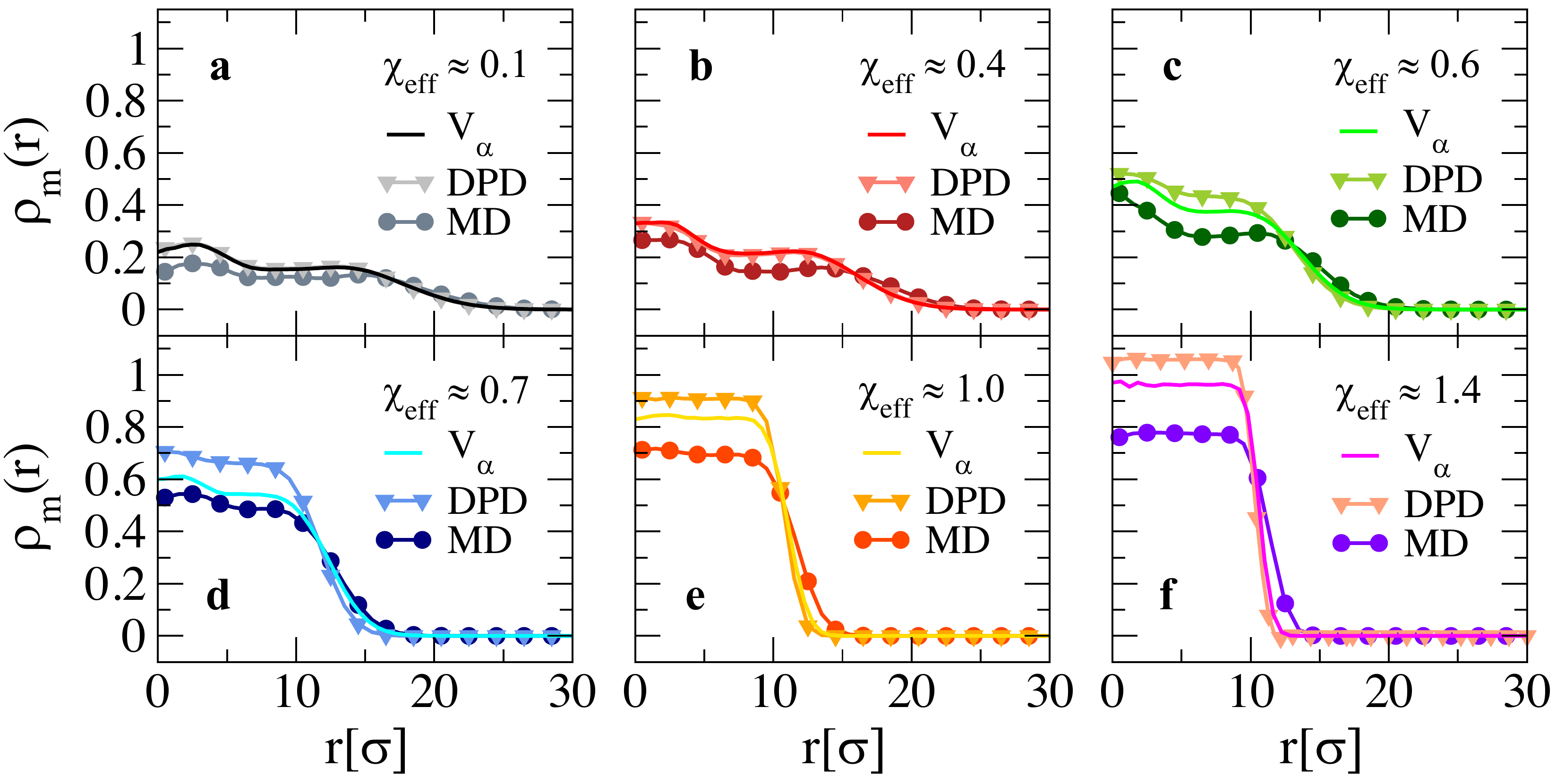}
\caption{\small \textbf{Density profiles for a loose microgel configuration across the VPT.} Monomer radial density profile $\rho_m(r)$ for a  $Z=25\sigma$ microgel as a function of the distance $r$ from its center of mass. Full lines refer to the implicit-solvent model, while symbols are used for MD (circles) and DPD (triangles) simulations. Each sub-panel refers to a different swelling state as in Fig. \protect\ref{fig:swellingcolori}(a).}
\label{fig:densprofmgel}
\end{figure}
We now discuss the structural features of the microgel at relatively large confinement, corresponding to the swelling curve in Fig.~\ref{fig:swellingcolori}(a). First, we show results for the density profile of the microgel in Fig.~\ref{fig:densprofmgel}, for several values of the swelling parameter across the VPT for both MD and DPD simulations. We find that, in general, both solvent models yield density profiles that are very similar to the implicit solvent case.  This is particularly true for the swollen states, where the typical core-corona structure of the microgels is clearly distinguishable. Under these conditions, DPD simulations are even more accurate than MD ones in reproducing the results of the implicit model. When $\chi_{\rm eff}$ increases and the microgel becomes more compact, the difference between the three models becomes more evident. Specifically, as the microgel collapses MD simulations produces lower density profiles in the core region with respect to the implicit-solvent case at the same $\chi_{\rm eff}$, while the DPD model generates more compact structures.

We notice that low density profiles exhibit a non-flat behavior in the inner core region of the microgel. These inhomogeneities, that are stronger for smaller microgels, can be removed out by averaging over independent topologies\cite{gnan2017}. Here we do not perform such an average because we aim to compare the behavior of a given microgel configuration with and without solvent. Beyond the VPT the oscillations are suppressed by the higher density, and hence the profiles are much flatter within the core.

\begin{figure}[t!]
\centering
\includegraphics[scale=0.445]{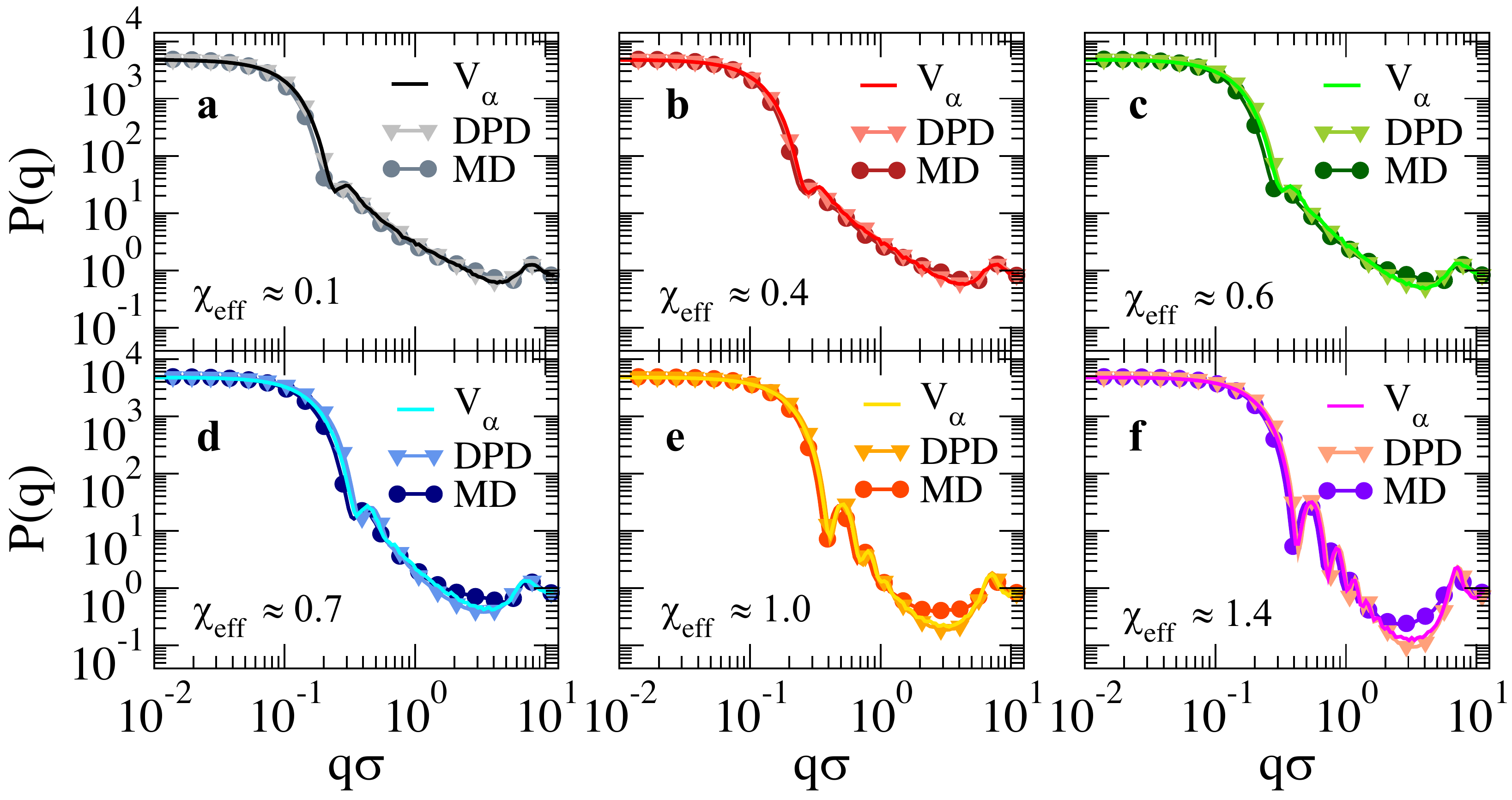}
\caption{\small \textbf{Microgel form factors for a loose microgel across the VPT.}  $P(q)$ as a function of $q\sigma$.  Full lines refer to the implicit-solvent model, while symbols are used for MD (circles) and DPD (triangles). Each sub-panel refers to a different swelling state according to Fig. \protect\ref{fig:swellingcolori}(a).}
\label{fig:formfactors}
\end{figure}

While density profiles provide real-space information on the microgel structure, they are not easily accessible in experiments, except for very recent super-resolution microscopy investigations\cite{scheffold,bergmann2018super}. Instead, they can be indirectly obtained from fitting the form factors to the fuzzy sphere model\cite{stieger2004thermoresponsive}. The form factors $P(q)$ can be measured by small angle neutron or x-ray scattering experiments. 
Thus, in contrast to density profiles, numerical $P(q)$ can be used to make a direct comparison with experiments, without having to rely on fits to specific models. Indeed, while the fuzzy-sphere model correctly describes the core-corona structure, it does not take into account the presence of dangling chains in the outer corona shell\cite{boon2017swelling,gnan2017}. 
We thus directly evaluate the form factors of the microgel across the VPT and present them in Fig.~\ref{fig:formfactors} as a function of wavevector $q$ for the same values of swelling parameters used in Fig.~\ref{fig:densprofmgel}. 
We find that the use of an explicit solvent does not considerably alter the form factors with respect to the implicit solvent case for all values of the swelling parameters.
As $\chi_{\rm eff}$ increases and the solvent quality decreases, $P(q)$ shows an increasing number of oscillations which become more and more pronounced. Furthermore, the position of the first peak, which is related to the microgel overall size, shifts to larger and larger wavevectors, indicating the shrinking of the microgel. 
However, a subtle difference is present between the two types of employed models: while DPD results are perfectly superimposed to the implicit solvent case for all $\chi_{\rm eff}$, the MD results are found to be always shifted to a slightly smaller $q$-value with respect to them. This is a reflection of the overall microgel size, which is a bit larger for MD explicit-solvent simulations with respect to DPD and implicit solvent, due to stronger excluded volume effects, as evident from Fig.~\ref{fig:cfr_swelling}.
We further notice that at relatively large wavevectors ($q\sigma \gtrsim 1$) the MD form factor systematically overestimates the DPD and implicit-solvent ones for intermediate and large values of $\chi_{\rm eff}$. However, all curves superimpose again at  $q\sigma\sim 7$, where a small peak is found, independently of the swelling parameter value. The latter corresponds to the monomer-monomer nearest-neighbour peak and is a feature associated to the excluded-volume interactions included in the bead-spring model for polymers and to the finite size of the simulated microgel. Indeed, for larger and larger microgel size, this peak would become more and more separated from the first one, allowing for a larger number of oscillations. In experiments, such a peak is not generally noticeable because of the soft intrinsic nature of the monomers. Thus, it is a limitation of the present modelling, which on very small length scales becomes inaccurate.

\begin{figure}[h!]
\centering
\includegraphics[scale=0.45]{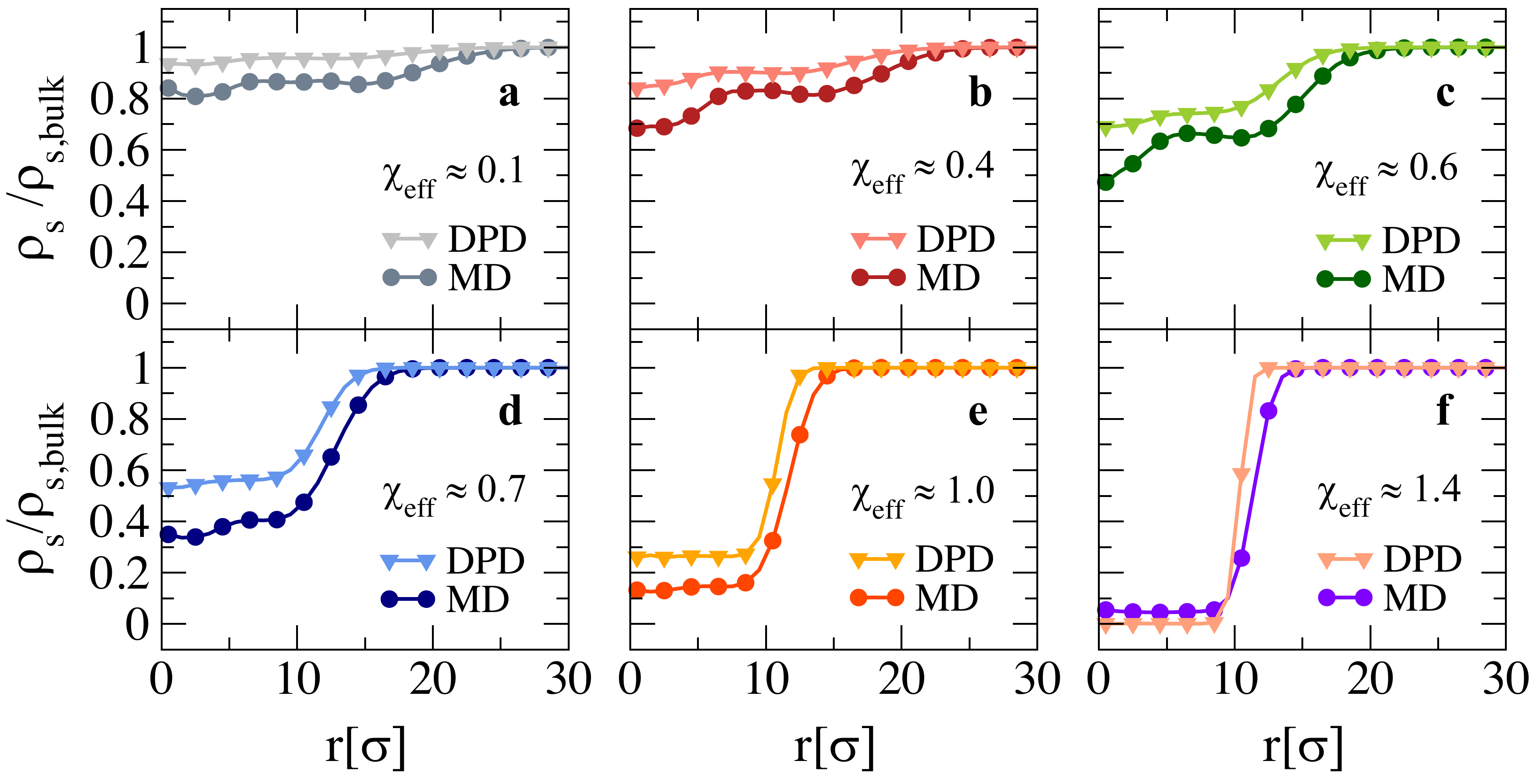}
\caption{\small \textbf{Solvent density profiles for a loose microgel configuration across the VPT.} We show the solvent density profile $\rho_s$ normalized with respect to the bulk solvent density $\rho_{s,bulk}$, as a function of the distance $r$ from the center of mass of the microgel. Circles and triangles refer to MD and DPD solvent, respectively. Each sub-panel refers to a different swelling state according to Fig. \protect\ref{fig:swellingcolori}(a).}
\label{densprofsolv}
\end{figure}

We now turn to analyze the solvent density profile $\rho_s$ inside the microgel.
The normalized profile $\rho_s/\rho_{s,{\rm bulk}}$, where $\rho_{s,{\rm bulk}}$ is the bulk solvent density, is shown in Fig. \ref{densprofsolv} as a function of the distance from the center of mass of the microgel. Clearly, the distribution reflects, as a mirrored image, the one of the microgel monomers. Indeed, when the core of the microgels becomes denser and denser, more and more solvent gets expelled. It is interesting to note that, beyond the VPT and except for the very collapsed states, a significant fraction of solvent is retained within the polymer network, even well inside the core region. At the VPT, which takes place at $\chi_{\rm eff}\sim 0.6$, the density of the solvent inside the core is larger than 50\% of the bulk value. 
Finally, we notice that there seems to be a consistent trend of the MD solvent to be more excluded from the network region with respect to the DPD results, again a feature associated to the larger excluded volume of the MD model. However, the two models yield qualitatively very similar results and reinforce the common view that microgels, despite their inhomogeneous structure and dense core region, retain $\gtrsim 90\%$ of water in their swollen configuration and still contain a large amount of water well beyond VPT, in qualitative agreement with the experiments results of Ref.~\cite{trappe}.

\subsection*{Results for a more confined microgel}
\begin{figure}[h!]
\centering
\includegraphics[scale=0.6]{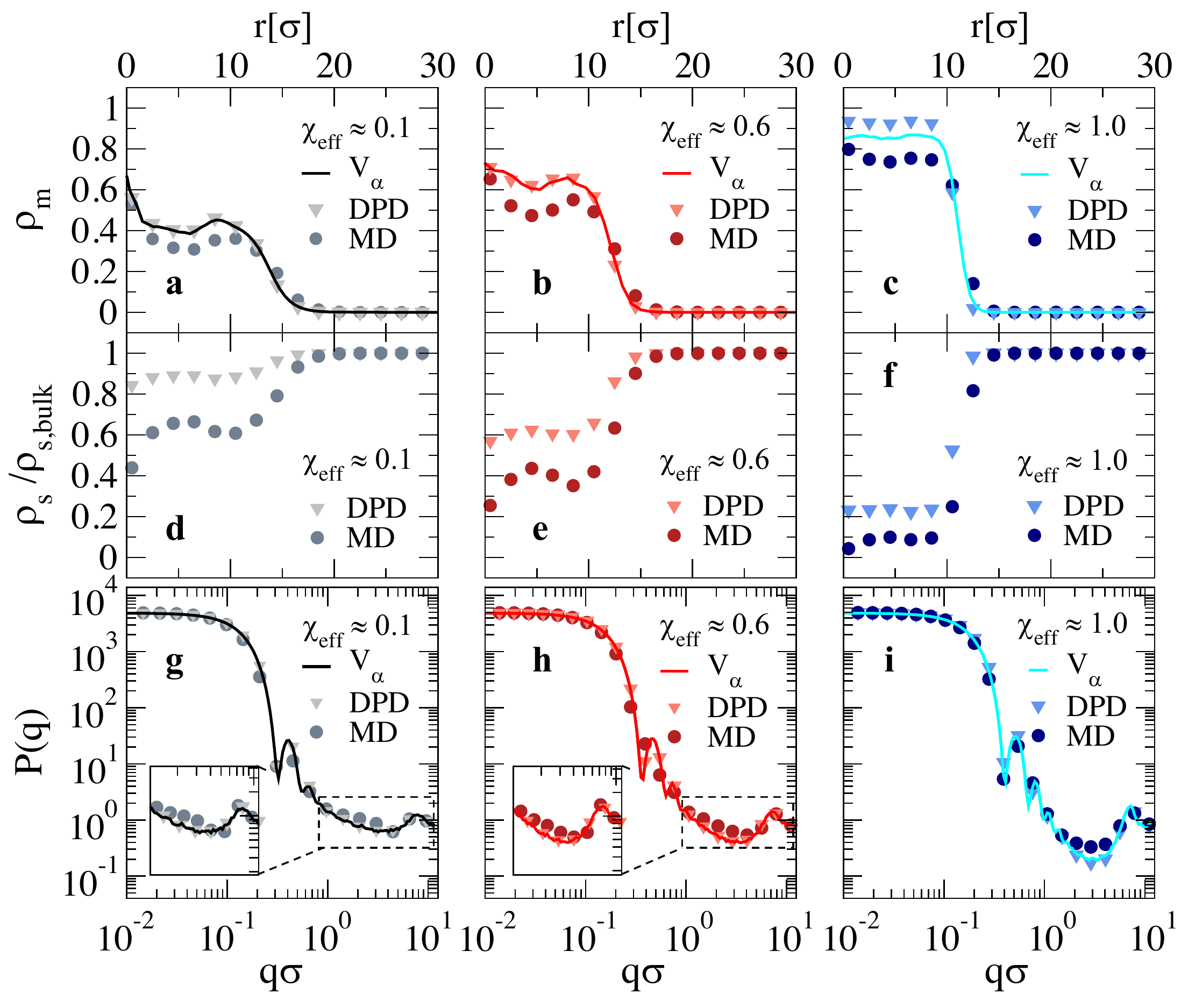}
\caption{\small \textbf{Microgel density profiles, solvent density profiles and form factors for a compact microgel across the VPT.} (a-c) microgel density profiles $\rho_m$ as a function of the distance $r$ from the center of mass of the microgel; (d-f) solvent density profiles $\rho_s$ normalized with respect to the solvent bulk density $\rho_{s,{\rm bulk}}$ as a function of $r$; (g-i) microgel form factors as a function of the wavenumber. Data are reported for a swollen state ($\chi_{\rm eff}=0.1$), a state close to the VPT ($\chi_{\rm eff}=0.6$) and a compact state ($\chi_{\rm eff}=1.0$). Full lines refer to the implicit solvent ($V_{\alpha}$), while symbols are used for DPD (triangles) and MD (circles). The insets in panels g and h show an enlargement of the high wavevector region where solvent-monomer excluded volume interactions induce an excess of signal for the MD data. } 
\label{fig:conf2}
\end{figure}
We now repeat the above structural analysis for a more compact microgel topology obtained with a smaller confining radius ($Z=15\sigma$), whose swelling curve was reported in Fig.\ref{fig:swellingcolori}(b). 

The density profiles of the microgel are reported in Fig.~\ref{fig:conf2}(a-c) for a few selected values of the swelling parameter and again for both MD and DPD explicit solvents. We find that the DPD model reproduces very well the implicit-solvent data, particularly for the more swollen conditions. When $\chi_{\rm eff}$ increases, the DPD monomer density in the core is slightly larger than for the implicit case. However, the corona profiles of the two microgel representations are identical. On the other hand, the MD solvent results underestimate the microgel density profile in the core and also display a different corona profile for all $\chi_{\rm eff}$. If compared to the findings for the looser microgel configuration (Figure~\ref{fig:densprofmgel}), the DPD solvent model behaves similarly for both types of networks and well reproduces the implicit model data in all cases. By contrast, the MD results present systematic differences with respect to the other two sets of data making the agreement not completely satisfactory. This is a consequence of the ``atomistic'' treatment of the solvent, which interacts via excluded volume with the polymer. Especially for compact microgels, when excluded volume becomes more and more relevant, these assumptions in the model may become unrealistic. Thus, while for looser networks both MD and DPD explicit solvents provide a good description of the microgel, for more compact microgels the DPD model has definitely the upper hand. This is also shown in the behavior of the solvent density profiles reported in Fig.~\ref{fig:conf2}(d-f). Again we find that the MD solvent is much more excluded from the interior of the microgels at all $\chi_{\rm eff}$. On the other hand, we see that, notwithstanding the relative higher compactness of this microgel, a significant amount of solvent remains inside the core in the swollen states, being roughly 60\% of its bulk value close to the VPT, in agreement with what found for the less confined microgel configuration and with experimental estimates\cite{trappe}.

The form factors, shown in Fig.~\ref{fig:conf2}(g-i), further confirm that DPD results are in good agreement with the implicit model ones. However, the MD outcomes display a clear shift in the peak position which is much more evident than for the looser configuration (see Fig.~\ref{fig:formfactors}). In addition, we observe an excess of signal, highlighted in the insets of Fig.~\ref{fig:conf2}(g,h), at $q\sigma \sim 3.0$ in swollen conditions, which is absent in the DPD and implicit solvent simulations. This difference occurs at a length that is roughly twice that of the monomer-monomer peak, thus being associated to monomers that are $\sim 2\sigma$ apart, \textit{i.e.} with a solvent particle in between them. Such a feature is smeared out at increasing $\chi_{\rm eff}$, when the microgel collapses and monomer-monomer interactions become dominant. We notice that the excess signal is not observed for the looser microgel as, at the same $\chi_{\rm eff}$ value, excluded volume interactions are far less important. Overall, this further shows that the MD model, while still acceptable for not too dense and open microgels, becomes more inaccurate for rather compact ones.

\subsection*{Collapse kinetics}
\begin{figure}[h!]
\centering
\includegraphics[scale=0.60]{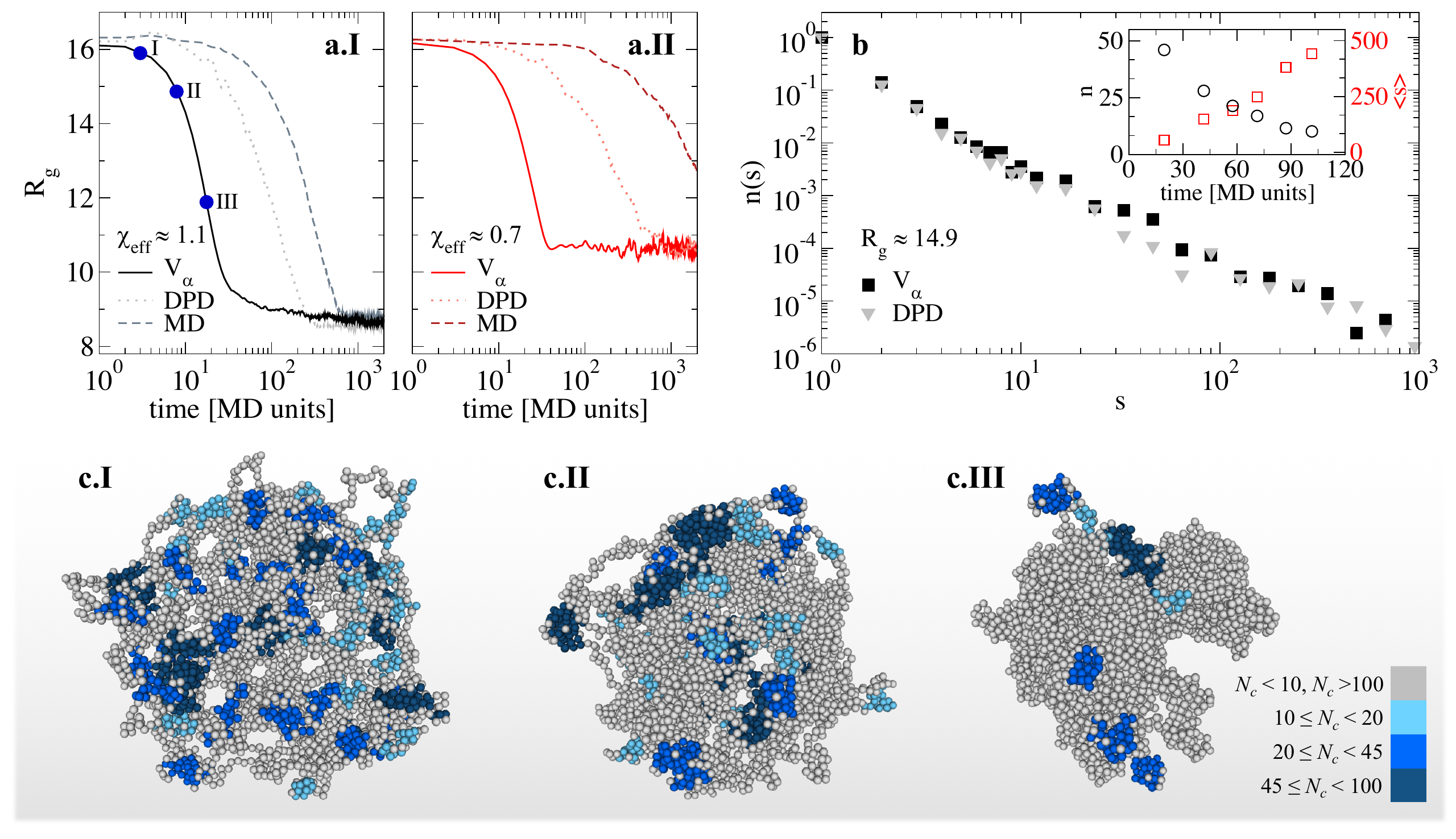}
\caption{\small \textbf{Collapse kinetics.} Radius of gyration $R_g$ as a function of time for a loose microgel ($Z=25\sigma$) for $\chi_{\rm eff}=1.1$  (a.I) and $0.7$ (a.II) for implicit ($V_{\alpha}$, full line), DPD (dotted lines) and MD solvents (dashed lines); (b) cluster size distribution $n(s)$ for $R_g=14.9$ (indicated as III in a.I) for implicit and DPD solvents. In order to improve statistics data are averaged over six different microgels configurations. The inset reports the number of clusters (black circles) and their average size (red squares) as a function of the collapsing time; (c.I-III) simulation snapshots for state points I-III (circles in a.I). Clusters are highlighted by different colors according to their size $N_c$ (as indicated in the color bar). Light grey monomers are either found in small clusters ($N_c <10$) or belong to the main network ($N_c>100$).
} 
\label{deswelling}
\end{figure}

After having established the explicit solvent models and having analyzed the properties of microgel and solvent particles in equilibrium for different values of the swelling parameters, we now turn our attention to the kinetics of collapse of the microgel in the presence of the solvent.
Employing the same approach adopted in Refs. \cite{reddy2006implicit,chang2001solvent,pham2008brownian,nikolov2018meso} for linear polymers, we start from a swollen microgel in a loose configuration and perform a sudden quench to a different state. In particular, we examine two final states whose value of $\chi_{\rm eff}$ correspond to an almost fully collapsed state ($\chi_{\rm eff}\sim 1.1$) and to a state close to the VPT ($\chi_{\rm eff}\sim 0.7$). We then assess whether the collapse transition is affected by the presence of the solvent by comparing the kinetics of the implicit-solvent model with that obtained using MD and DPD ones. 
Figure \ref{deswelling}(a.I-II) shows the time evolution of the radius of gyration of the microgel for the three different types of simulations at two different $\chi_{\rm eff}$. In all cases the curves reach at long time the same value of $R_g$ but, in these simulation conditions, the time taken to equilibrate is different, being faster in implicit solvent simulations compared to those of DPD and MD (the slowest). All curves display a sharp one-step collapse with no trapping phenomena in metastable states. This is qualitatively in agreement with experiments in which microgels with a similar core-corona structure to ours are subjected to an abrupt temperature jump from low (swollen state) to high temperature (globular state) \cite{seiffert2012impact}.

In order to highlight the role of the solvent, we perform a cluster analysis to identify how the microgel structure evolves during the collapse. To this aim, we detect clusters of non-bonded monomers (see Methods), and calculate their size distribution for state points having the same $R_g$ but simulated with different models. Remarkably, we find the same cluster distribution for both implicit and DPD solvent, indicating that the solvent plays no significant role on the folding dynamics of the microgel, as shown in Fig.~\ref{deswelling}(b). To visualize the restructuring of the microgel following the instantaneous decrease in the solvent quality, snapshots of the microgel are reported in Fig.~\ref{deswelling}(c.I-III) for three different times. The microgel, while shrinking, first reorganizes by grouping monomers into small clusters (panel c.I). Each cluster is connected to the others via single or multiple links so that the structure, at an intermediate shrinking stage, displays a large number of holes and becomes increasingly inhomogeneous. As the shrinking proceeds, the clusters start to merge, becoming larger and larger in size and joining the main network (panels c.II-III). Finally, at long times, all non-bonded monomers are connected and only a single cluster is left. The decrease in the number of clusters as well as the increase of their average size as functions of the collapsing time are evidenced in the inset of Fig.~\ref{deswelling}(b).
We stress that the same kinetic pattern is also found for the implicit model simulations and for the more confined microgel (not shown). In addition, a recent work~\cite{moreno2018} has reported a very accurate analysis of the collapse dynamics of microgels which is found to be in qualitative agreement with our findings. These results strongly indicate that the solvent plays a minor influence on the structure of the microgel during the collapse transition. Indeed, at each swelling stage, the microgel has a similar structure regardless of the solvent employed, suggesting that deswelling occurs via the same sequence of transient states. It would be interesting to compare these findings with more accurate solvent treatments such as Multi-Particle-Collision-Dynamics simulations\cite{ghavami2017solvent,kobayashi2014structure}.

\section*{Discussion}

The tunable swelling of the microgel particles has been, since their discovery, one of the most relevant features of these colloids. Indeed, the opportunity to tune the particle volume fraction without changing their number density, but only the temperature, is a formidable advantage for experimental investigations. However, this poses a computational challenge in choosing a suitable model that best describes their swelling-deswelling transition. The recent assembly of realistic microgel networks in Ref.\cite{gnan2017} correctly reproduces experimental density profiles and form factors through an implicit solvent treatment. However, the inclusion of the solvent grants additional information, such as the uptake of solvent within the polymer network or surface tension effects. For these reasons, in this work we have compared the implicit solvent results to explicit solvent ones by employing two common approaches to simulations that allows for an atomistic and a coarse-grained approach, namely MD and DPD.
We found that we can reproduce the implicit solvent swelling behavior by tuning the monomer-solvent interaction potentials after having adjusted the solvent density. This stems from the fact that, when the solvent is treated explicitly, the external pressure exerted by the solvent needs to be adjusted. In DPD simulations, the same effect can be obtained by regulating the cut-off radius.

We considered two microgels differing in the degree of compactness, which can be obtained by different synthesis protocol\cite{habicht2014non} and/or by varying the number of crosslinkers. We found that, particularly when the network is denser, excluded volume interactions play a relevant role in the description of the microgel. Indeed, in the full MD simulations an additional peak in the structure appears at small length scales. At the same time, the internal density profile of the microgel is also affected, resulting in a less dense core and a modified corona behavior, which is more significant in the collapsed state.
Despite reducing the size of the solvent may solve excluded volume issues, doing so would dramatically increase the number of particles required to observe the same swelling behavior, as the box size is fixed by the dimensions of the microgel particle.

By contrast, DPD results better describe the implicit model ones for both microgel density profiles and form factors, at all swelling conditions. Furthermore, the DPD model reproduces the behaviour of the radius of gyration of the implicit model at different swelling conditions in an almost quantitative fashion. We have also investigated to what extent the solvent penetrates into the microgel, and we found that in the MD simulations much less solvent is present in the interior of the network, whereas DPD results seem more realistic in comparison to experimental estimates. 
Indeed, we find that, in the swollen state, the network is completely hydrated, retaining more than 90\% of the solvent (with respect to the bulk density) in the core of the microgel. Even above the VPT the microgel contains a large fraction of solvent, which is finally excluded only at very large $\chi_{\rm eff}\gtrsim 1.4$, amounting to temperatures $\gtrsim 60^{\circ}$C according to the mapping established in Ref.~\cite{gnan2017} for PNIPAM microgels.

We also examined the collapse kinetics and assessed how the presence of the solvent affects it. We observed that, in the conditions we performed simulations, a slowing down of the collapse dynamics occurring for the more structured solvent (MD simulations) and to a smaller extent for the coarse-grained solvent (DPD simulations) with respect to the implicit simulations.
However, we also found that the system, when compared at the same swelling degree (quantified by the radius of gyration of the microgel), always presents a similar structure, regardless of the model. In particular, at first the network becomes rather inhomogenous, with regions where monomers have clustered together and empty regions. Later on, the clusters merge together and become larger and larger, until the collapse is complete and the microgel is essentially a fully folded network. Such transient behavior, featured by the appearing of crumples, has also been observed previously in simulations\cite{chang2001solvent,pham2008brownian,nikolov2018mesoscale}. The similarity between these results with those found for an implicit solvent treatment suggests that hydrodynamic interactions do not play a major role in the swelling-deswelling transition, which is instead mainly controlled by the quality of polymer-solvent interactions.

In summary, in this work we have established that DPD simulations with a coarse-grained solvent constitute the most suitable method to include explicitly a generic solvent in the simulation of a microgel colloid.
Even though a partially satisfactory description can be also obtained with the use of an MD solvent, the atomistic description allows for the presence of significant excluded volume interactions that brings unphysical features in the model. On the other hand, DPD simulations do show a full agreement with the implicit model and provides a realistic description of the solvent arrangement within the network. Thus, our model of realistically assembled microgels in DPD explicit solvent opens up the possibility to tackle those phenomena where the physical presence of the solvent is crucial. In particular, our model may serve as a starting point to numerically investigate the so-called “Mickering" emulsions\cite{schmidt2011influence} and in the fascinating case of microgels at  fluid-fluid interfaces\cite{isa2017two,geisel2012unraveling,scheidegger2017compression,
brugger2010interfacial,rumyantsev2016polymer}. Finally, a realistic description of how the solvent is trapped within the polymer network may lead to future advances in the field of drug delivery and controlled-release, and can provide further insights into its mechanism\cite{li2004numerical,bysell2011microgels}. 

\section*{Methods}
\small
\textbf{Microgel assembly.} The starting configuration of a microgel particle is prepared as in Ref.~\cite{gnan2017}. First, we produce a fully-bonded disordered network by self-assembling a binary mixture of bi- and tetravalent patchy particles with an applied spherical confinement. Once the network has assembled, we replace the patchy interactions with the classical bead-spring model for polymers\cite{grest1986molecular}, in which bonded monomers interact via the sum of a Weeks-Chandler-Andersen \cite{wca1971} (WCA), $V_{WCA}(r)$, and a Finite-Extensible-Nonlinear-Elastic\cite{bernabei2011chain,soddemann2001generic} (FENE), $V_{FENE}(r)$, potentials:
\begin{equation}\label{wca}
V_{WCA}(r)=\begin{cases}
    4\epsilon\left[\left(\frac{\sigma}{r}\right)^{12}-\left(\frac{\sigma}{r}\right)^{6}\right]+\epsilon & \text{if $r \le 2^{\frac{1}{6}}\sigma$}\\
    0 & \text{otherwise}
  \end{cases}
  \\;\ \ \ \ \ \ \  V_{FENE}(r)=-\epsilon k_FR_0^2\ln(1-(\frac{r}{R_0\sigma})^2)      \text{ if $r < R_0\sigma$}
\end{equation}
with $k_F=15$ an adimensional spring constant and $R_0=1.5$ the maximum extension value of the bond. Non-bonded monomers only experience a repulsive WCA potential. Regarding units, lengths are given in units of $\sigma$, which corresponds to the diameter of a monomer of unit mass $m$, energy in units of $\epsilon$ and time in units of $\sqrt{m\sigma^2 /\epsilon}$.

\noindent
In this work, we build networks of $N\sim 5000$ monomers confined within a sphere of two different radii $Z$, namely $Z=15\,\sigma$ and $Z=25\,\sigma$. The change in $Z$ allows to vary the topology of the network, which becomes more compact for small $Z$ and looser (and with more dangling ends) for larger $Z$~\cite{gnan2017,rovigatti2018internal}. 
The number of crosslinks is fixed to $3.2\%$ of the total number of monomers.

\noindent \textbf{Implicit solvent.} The implicit solvent is modeled through the addition of an attractive potential $V_\alpha$ that acts between all monomers, either bonded or non-bonded, of the microgel \cite{soddemann2001generic,verso2015simulation}:
\begin{equation}\label{alpha}
V_{\alpha}(r)=\begin{cases}
    -\epsilon\alpha & \text{if $r \le 2^{\frac{1}{6}}\sigma$}\\
    \frac{1}{2}\alpha\epsilon\left[\cos\left(\delta\left(\frac{r}{\sigma}\right)^2+\beta\right)-1\right] & \text{if $2^{\frac{1}{6}}<r\le R_0\sigma$}\\
    0. & \text{otherwise}
  \end{cases}
\end{equation}
where $\delta=\pi(2.25-2^{1/3})^{-1}$ and $\beta=2\pi-2.25\delta$~\cite{soddemann2001generic}. 
The potential is modulated by the parameter $\alpha$, which controls the solvophobicity of the monomers and plays the role of an effective temperature. For $\alpha=0$, monomers do not experience any attraction and good solvent conditions are reproduced while, by increasing $\alpha$ up to 1.5, the monomers become fully attractive, mimicking bad solvent conditions. Previous analysis has shown that the volume phase transition takes place at $\alpha\sim 0.6$~\cite{gnan2017}. MD simulations are performed at constant reduced temperature $T^*=k_BT/\epsilon=1$ (where $k_B$ is the Boltzmann constant) using the Nos\`e-Hoover thermostat and a timestep $t^*=0.002$.

\noindent \textbf{Adding an explicit solvent in MD simulations.} We take a configuration of the microgel assembled as described above and perform MD simulations in the presence of a varying number of additional spheres that mimic the solvent particles which, for efficiency reasons, have also a diameter $\sigma$. The number of solvent particles varies between $2.5\times 10^5$ and $3\times 10^5$ in a simulation box of size $L=70\,\sigma$, yielding solvent number densities $0.729 \le  \rho_s \le 0.845$, for which the LJ solvent is in the fluid regime. Lower densities would bring the LJ solvent to phase separate, while higher $\rho_s$ would lead to a crystallization of the solvent particles.  
All MD simulations with explicit solvent are performed with the LAMMPS simulation package~\cite{plimpton1995fast} at $T^*=1$ making use of the Nos\`e-Hoover thermostat and a timestep $t^*=0.002$.  The center of mass of the microgel is fixed in the center of the simulation box. To model \textit{solvent-solvent} interactions we use a Lennard-Jones (LJ) potential, $V_{LJ}(r)= 4\epsilon\left[\left(\frac{\sigma}{r}\right)^{12}-\left(\frac{\sigma}{r}\right)^{6}\right]$. 
Here $\epsilon$ is the same as the one used in the WCA of monomer-monomer interactions.

\noindent The choice of the \textit{monomer-solvent} (ms) interactions is crucial in order to implement the solvophobic effect, giving rise to the volume phase transition of the microgel. In this respect, we test different approaches.
Our first choice is to employ again the LJ potential in which its depth $\epsilon_{\rm ms}$ is varied,
so that the attractive contribution can be tuned. A weaker attraction would thus give rise to a more repulsive monomer-solvent interaction that should cause the shrinking of the microgel. However, as explained by analyzing the swelling curves in the Results section, a decrease of this parameter causes the repulsive barrier to be less efficient, giving rise to unphysical consequences on the swelling behavior. To fix this problem, we consider a $\lambda$-dependent Lennard-Jones potential~\cite{rovigatti2016soft}, $V_{\lambda}(r)$, defined as:
\begin{equation}\label{lambda}
V_{\lambda}(r)=\begin{cases}
    V_{WCA}-\epsilon\lambda & \text{if $r \le 2^{\frac{1}{6}}\sigma$}\\
    4\epsilon\lambda\left[\left(\frac{\sigma}{r}\right)^{12}-\left(\frac{\sigma}{r}\right)^{6}\right] & \text{otherwise}
  \end{cases}
\end{equation}
where $\epsilon$ is the same as for the monomer-monomer (Eq.~\eqref{alpha}) and LJ solvent-solvent interactions, while $\lambda$ plays the role of an inverse temperature (analogue to the inverse of $\alpha$ in the implicit solvent model). Indeed, for large values of $\lambda$ there is an attractive contribution between a monomer and a solvent particle, mimicking good solvent conditions, while for $\lambda=0$, the WCA potential is recovered and monomer-solvent interactions are purely repulsive. The potential is truncated and shifted at $2.5\sigma$. The advantage of using such a potential with respect to the simple LJ interactions is that it allows to alter the monomer-solvent interactions, and thus the “quality" of the solvent, without affecting the excluded-volume part; this remains encoded in the $V_{WCA}$ term and it does not depend on $\lambda$.

\noindent \textbf{Adding an explicit solvent in DPD simulations.}
DPD is a mesoscale simulation technique\cite{groot1997dissipative,keaveny2005comparative} that treats solvent particles as coarse-grained beads and is able to describe hydrodynamic interactions through a momentum-conserving thermostat. 
In DPD simulations, particles $i$ and $j$ interact by three pairwise additive forces: a conservative force $\vec{F}^C_{ij}$, a dissipative force $\vec{F}^D_{ij}$ and a random force $\vec{F}^R_{ij}$ where
\begin{equation}\label{dpd_fc}
\vec{F}^C_{ij}=\begin{cases}
    a_{ij}(1-r_{ij}/R_c)\hat{r}_{ij} & \text{if $r_{ij}<R_c$},\\
    0 & \text{otherwise}
  \end{cases}; \ \ \ \ \ \vec{F}^D_{ij}=-\gamma w^D(r_{ij})(\hat{r}_{ij} \cdot \vec{v}_{ij}) \hat{r}_{ij}; \ \  \ \ \ \vec{F}^R_{ij}=\sigma_R w^R(r_{ij}) \theta (\Delta t)^{-1/2}\hat{r}_{ij}.
\end{equation}
Here $\vec{r}_{ij}=\vec{r}_i-\vec{r}_j$ with $\vec{r}_{i}$ the position of particle $i$, $r_{ij} = |\vec{r}_{ij}|$, $\hat{r}_{ij} = \vec{r}_{ij} / r_{ij}$, $\vec{v}_{ij}=\vec{v}_i-\vec{v}_j$ with $\vec{v}_{i}$ the velocity of particle $i$,  $w^D(r_{ij})$ and $w^R(r_{ij})$ are weight functions, $\theta$ is a Gaussian random number with zero mean and unit variance and $\gamma$ is the friction coefficient (here $\gamma=4.0$); to ensure that the correct Boltzmann distribution is achieved at equilibrium, $w^D(r_{ij})=[w^R(r_{ij})]^2$ and $\sigma^2_R=2\gamma k_BT$. The interaction region for the dissipative force is defined in the same way as for the conservative force, \textit{i.e.} $w^D(r_{ij})=1-r_{ij}/R_c$. We refer the reader to Ref.~\cite{groot1997dissipative} for more details on the DPD simulation technique.

\noindent
Although the DPD protocol is often applied to all the species that are present in the simulation, here we make use of DPD only for the solvent-solvent and the solvent-monomer interactions, while keeping the microgel model unaltered. This hybrid technique allows to simulate a “fast" solvent by retaining important features of the microgel particle, such as the topology and the excluded volume interactions among monomers, already investigated in the implicit solvent case \cite{gnan2017}. In DPD simulations, we fix the solvent number density at $\rho=0.73$, using $N=2.5\times10^5$ particles in a simulation box of side $70\sigma$, and we tune the interaction parameters and the radius of the solvent beads until the swelling curve of the implicit solvent model is reproduced, in this case at $r_c=1.75\sigma$. The same curve may be found by using different combination of these parameters in the limit in which the size of the solvent bead is comparable with that of the microgel monomer.
All DPD simulations are also performed with LAMMPS at $T^*=1.0$ using the velocity-Verlet algorithm to integrate the equations of motion; the DPD thermostat is applied to the solvent particles only. Moreover, the center of mass of the microgel is fixed in the center of the simulation box as in MD simulations.
\noindent
The monomer-solvent interaction parameters $a_{ij}^{\rm ms}$, that for simplicity we call $a_{\rm ms}$, plays the role of an effective temperature and, depending on its value, controls the volume phase transition of the polymer network. The repulsion coefficient for solvent-solvent interactions is fixed at $a_{ij}^{\rm ss}\sigma=a_{\rm ss}\sigma=25\epsilon$.

\noindent \textbf{Rescaling of swelling curves.} The swelling degree of the microgel is expressed via the ratio between the absolute value of the gyration radius and its maximum value, obtained in good solvent conditions. The gyration radius is computed as $R_g=[N^{-1}\sum_{i}^N (\vec{r}_i-\vec{r}_{CM})^2]^{1/2}$, where $\vec{r}_{CM}$ indicates the position of the microgel center of mass. Since each model depends on a different ``swelling parameter'', that we call generically $\chi_{\rm eff}$, we scale all curves onto the implicit model one, using $\alpha$ as the reference swelling parameter. 
For those explicit solvent models where a small value of the swelling parameter corresponds to a collapsed state of the microgel, \textit{i.e.} $V_{LJ}$ and $V_{\lambda}$, the scale has to be inverted. In order to properly rescale the $x$ axes onto each other for two curves $A$ and $B$, we consider two points on the first ($x_1^A$ and $x_2^A$) and on the second curve ($x_1^B$ and $x_2^B$), respectively. The rescaled $x$-coordinate is calculated using the following relationship: $x_{new}=\left(x-\langle x^A \rangle\right)\Delta x^B/\Delta x^A  + \langle x^B \rangle$, where $\langle x^i \rangle=0.5(x_1^i+x_2^i)$ and $\Delta x^i= x_1^i-x_2^i$ with $i=A,B$. 

\noindent \textbf{Form factors.} The microgel form factor $P(q)$ is calculated as $P(q)=\frac{1}{N}\sum_{ij} \langle \exp{(-i\vec{q} \cdot \vec{r}_{ij}) \rangle}$ where the angular brackets indicate an average over different equilibrium configurations of the same microgel and over different orientations of the wavevector $\vec{q}$.

\noindent \textbf{Cluster analysis in the kinetics of deswelling.}  To investigate the structural changes during the transient kinetics of deswelling, we define clusters within the microgel that are formed by non-bonded monomers only: two such monomers belong to the same cluster when their distance is smaller than $1.2\sigma$, which roughly corresponds to the first peak of the radial distribution function $g(r)$. Such a restrictive choice of the cut-off distance, for which lots of monomers in the first coordination shell are effectively disregarded, makes it possible to generate a complete distribution of cluster sizes. Indeed, in our microgel configuration, all the monomers are connected into a single cluster by definition, in force of the intra-polymer bonds. Thus we cannot resort to common values of the cut-off, such as the minimum of the $g(r)$, or to more refined cluster algorithms based on nearest neighbours~\cite{van2012parameter}, because they would yield only few large clusters in the system, making it difficult to compare the distributions obtained with different methods. The distribution of clusters of size $s$, $n(s)$, presented in Fig. \ref{deswelling}, is calculated by averaging over six independent configurations of the microgel.

\normalsize
\section*{Acknowledgements}
We acknowledge support from the European Research Council (ERC Consolidator Grant 681597, MIMIC).

\section*{Author contributions statement}
FC, NG, LR and EZ performed simulations, analyzed results and wrote the paper.

\section*{Additional information}
The authors declare no competing interests.

\bibliography{mybib}

\end{document}